# Edges fractal approach in graphene – defects density gain

I. Janowska,[a]* M. Lafjah,[a] V. Papaefthymiou,[a] S. Pronkin,[a] C. Ulhaq-Bouillet[b]

[a]Institut de Chimie et Procédés pour l'Énergie, l'Environnement et la Santé (ICPEES), CNRS UMR 7515-University of Strasbourg, France

[b]Institut de Physique et Chimie des Matériaux de Strasbourg (IPCMS), CNRS UMR 7504-University of Strasbourg, France

*Corresponding author: janowskai@unistra.fr, tel: 33 (0)368852633

**To optimize the technological development together with energy gain, the crucial materials needs to be tailored in a smart way to benefit a maximum from the property-structure relation. Herein we propose the way to increase a defect density in graphene moving from 1D linear edges to fractal edges, also highlight the fractal nature of graphene structures them-self. The tentative oxidation and hydrogenation catalytic etching of few layer graphene results in structures with edges fractals, jaggy edges at graphene periphery or in newly formed etched holes.**

The sustainable development emerging from accelerating societal requirements is a driving force for the recent progress in green technologies and crucial (nano)materials. To face up the economical and environmental issues a smart choice of crucial materials including their abundance, price, production and appropriate tailoring is necessary.

Due to diver, often baronial and combined properties, the important materials of choice are graphitic materials and especially 2D graphene or few layer graphene and their composites. Although their intrinsic features are mostly related to the C=C conjugated honeycomb lattice, their efficient use calls for application dependent tailoring of their chemical/electronic/geometry and are often related to the presence of defects. Different structural defects (Stone-Wales, vacancy, point, linear, edges), oxygen, hydrogen functionalities or doped lattice heteroatoms induce a positive or penalizing effect and are intended to be removed or introduced, consequently. The intentional introduction of defects aims often to improve or even provide the reactivity (i.e. metal-free catalyst), hydrophylicity, (opto) electronic, magnetic or sensing properties among others [1]. The important defects are graphene edges, the linear defects usually introduced by a catalytic etching (cutting) of graphene lattice. Since the etching of graphite is known for dozen of years, its recent revival concerns mostly the hydrogenation etching with transition metal NPs (Ni, Co, Fe) to reach the nanoribbons or dots possessing well defined "arm-chair" or "zig-zag" edges conformation [2,3]. The preferential etching direction is determined by graphene lattice crystallography and no matter the direction is, the most reported work focus on straight line channels formation if not bearing in mind the arm-chair" or "zig-zag conformation at atomic level related to $sp^2$ hybridized C atom geometry.

The synthesized and tailored graphitic materials belong typically to Euclidean geometry (1D tubes or ribbons, 0D fullerens or dots, 2D graphenes, 3D graphites), meanwhile an expanded in 2D space the hexagonal structure in perfect graphene can already be considered as a hierarchical multifractal [4]. Only one group reported at presence the catalytic in-*situ* fractal etching of CVD-graphene [5], and it is much worthy to urgently take care for non-integer dimension of graphenes. A defined by Mandelbrot "fractals" are indeed omnipresent in natural matter geometry, physicochemical phenomena, mathematical analysis methods, and fractal design is more and more explored. The scarce reports link to graphene. Very recently, a fractal quantum Hall effect in graphene-BN heterostructure and structure-due tenfold enhancement of photocurrent generation in fractal metasurface graphene designed by e-beam lithography, were observed [6, 7]. The fractal approach was also applied for the formation of conductive self-assembled fractal branched patterns, with reduced percolation threshold for a given substrate surface [8].

Here, we highlight the fractal feature of graphene and exemplary possibilities offered by graphene structure in term of its fractal geometry pattering but primary we focus on the patterning of the graphene edges. Essentially, we propose a new look on the graphene defects (edges), which can have space expanding fractal dimension (and can have mixed zig-zag and arm-chair conformations at atomic level). We claim that the density of active sites in graphene can be strongly enhanced not only by increasing the edges number but increasing it in nonlinear manner to reach, "jaggy", fractal edges. The concept reflects a known fractal expansion of 1D line in 2D plane, with exemplary illustrated in fig. 1a. Such successive expansion (from I to III) increases the length of the line at the distance from point A to B. Let's imagine now these lines are the edges of graphene, in other words: active sites. The number of active sites does not increase by multiplication (or elongation) of straight edges, as it is a case of most of the etched channels, but by increasing their density in 2D plane. To achieve the length of edge III (or the same active sites number) by simple elongation of edge I the latter needs to be elongated almost twice. The examples of such pattering in nature are common including the plants' leaves (fig. 1b).

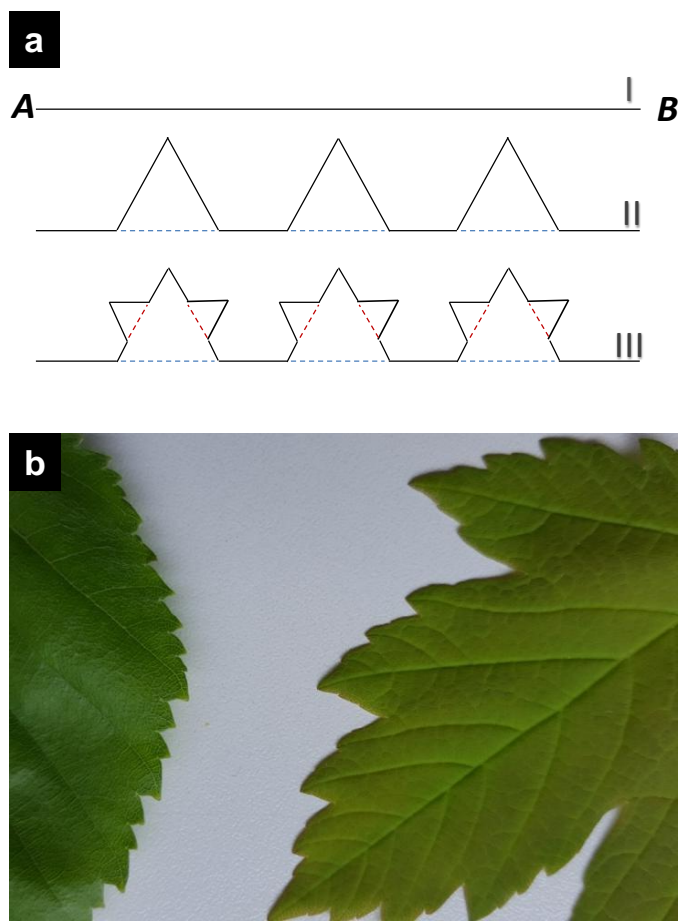

Fig. 1 a) Schema of exemplary fractal expansion of a 1D line, b) optical photo of trees' leaves with eye visible jaggy edges (left, right) and fractal edges/structure (right).

While the reported etching of graphene mostly looks for straight line channeling, any asymmetry of the etched lines, rarely discussed, is seen by us as benefic for active site density enhancement. We have already observed earlier some partially asymmetric, "step-like", edges in a case of $FeN_x$ NPs assisted etching, were the "creep" movement of the faceted active $FeN_x$ NP occurred [3]. The intended and controlled expansion of the edges through the right experimental approach would be however highly suitable.

At present we show the results of tentative graphene etching, which proceeds in a way to provide the graphene structures without the pronounced channels but with asymmetric, strongly "jagged" edges. For this purpose, few layer graphene (FLG) flakes obtained previously by liquid exfoliation method, of lateral size ~3 µm and decorated with $Fe_{3-x}O_4$ (SI) was submitted to the oxidative treatment with air flow under atmosphere pressure at low temperature (300°C) for 0.5h. Next to this, the temperature was increased to 800°C under Ar and hydrogenation treatment with $H_2$/Ar flow was applied. Fig. 2 a shows TEM micrographs of the Fe-based/FLG structures after the performed treatment. For the reason of comparison, the results from the etching running under "standard" hydrogenation conditions at 800°C with the same catalytic system are also presented in fig. 2 b. The difference between the samples submitted to oxygen assisted and purely hydrogen assisted treatments is clear. While in the case of "standard" hydrogenation the straight line channels with the active NPs at their ends are formed in the FLG flakes under well defined angles directions to periphery edges (often perpendicularly) (fig. 2b), the addition of pre-oxidation process results in the flakes sheets with "jaggy" edges and the channels are *quasi* absent (fig. 2a).

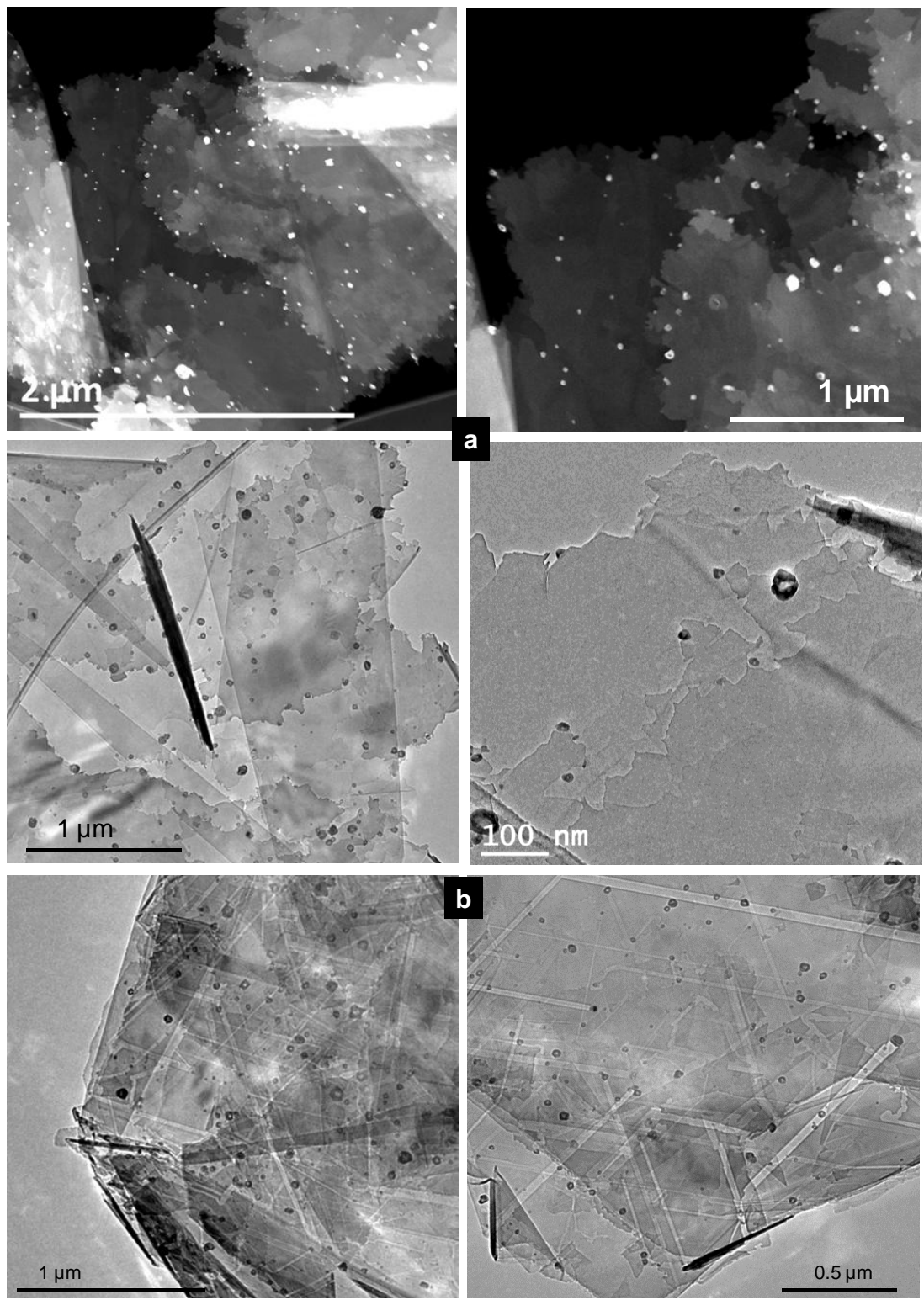

Fig. 2 TEM micrographs of a) FLG with jagged edges obtained after oxidation, hydrogenation treatment, b) FLG with straight edges etched by hydrogenation

At the moment it is difficult to make a final statement about the impact of the atomic oxygen potentially produced from the dissociative chemisorptions of molecular oxygen on NPs surface and of the NPs themselves. It can be found indeed in the literature that in the case of iron oxide, a dissociation and recombination of oxygen can take place at the temperature even lower then oxidation of NPs. The recent work concerning the investigation of oxidation of graphene supported on Ir(111) also claimed the role of atomic oxygen, which production was facilitated by Ir(111) at the graphene edges [9]. Those oxidized edges were however only slightly "jagged", rather wavy, at lower scale and degree that the ones presented here. Other thing is that we did not observe any oxidative channeling of FLG $Fe_{3-x}O_4$ NPs after oxygen treatment at 300°C. We suppose then that the strongly asymmetric "jaggy" edges obtained with $O_2$ + $H_2$/Ar treatments are the effect of both the oxidation through atomic oxygen and hydrogenation process. The lack of numerous visible typical long straight channels after the subsequent hydrogenation can be explained by strongly oxidized edges, excessive coalescence of NPs and rapid etching. All together result in fast etching and sometimes in total gasification of the etched path. If then the NPs are larger than the thickness of FLG, the flakes cut in small pieces can be observed from time to time (TEM, SI). This confirms the general statement of the etching proceeding easier at more defected/oxygen rich carbon, the reason also why the channeling in general initiates at graphene edges. The primary importance is however the fact that most of the herein etching directions are not well oriented by crystallographic lattice of graphene, which is breached, and NPs change often their trajectory adopting the creep movement due to modified interactions with oxidized edges. Contrary then to several works investigating the oxidation of graphene (graphite) as a thermal/air stability drawback; we see the controlled oxidation as support for defect density rich tailoring of graphene planar structures.

Since the edges are "jugged" at various degrees we can consider them as fractal edges, according to description of schema 1, in which the equilateral triangular extension may be replaced by less symmetric or others geometry analogue such as for example trapeze.

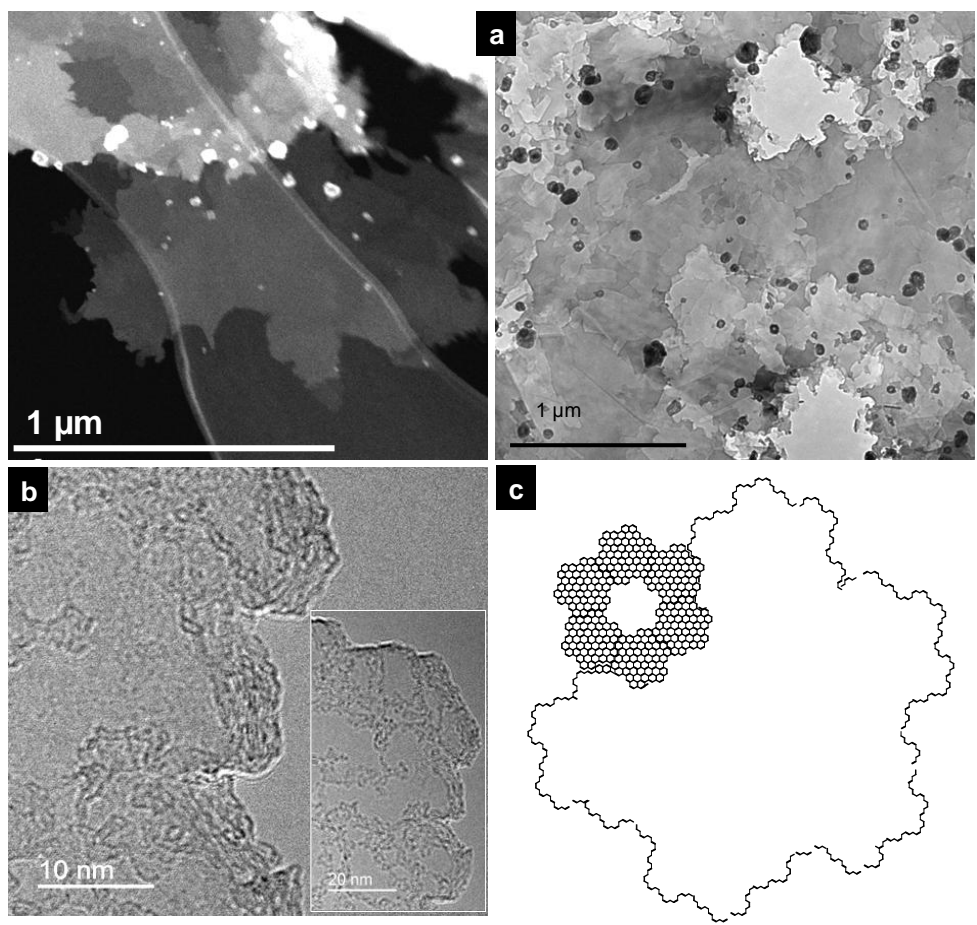

Fig. 3 a) TEM micrographs of FLG with periphery jagged edges and formed holes, also with jagged edges obtained after oxidation, hydrogenation treatment b) representative TEM micrograph of jagged FLG edge, c) schema of benzenoid with rotary symmetry of n = 4.

Apart from the periphery jaggy edges, the etched large holes, also possessing jagged edges have been observed (fig. 3a). They formation is benefic for additional enhancement of the edges-to-plane ratio and so increasing of defects density, as well as for introducing the porosity in the bulk material. Fig. 3b shows representative TEM micrographs of the edges.

The fractal edges form part of "surface fractal", where the surface corresponds to "surface of edges" as analog to other surfaces, for instance, the chemically active surface of metal NPs. In this context and from a chemistry point of view, the well crystallized graphene planar sheet are rather a platform for defects active sites, especially in few layer graphene, similarly to bulk of NPs (obviously the mutual interactions across C=C conjugated planar lattice and defects exist). Due to however the much higher accessibility of the graphene plane (monolayer) towards reactants compared to the NPs bulk and the fact that the word "surface "is firmly linked to 2D planar graphene geometry, the formulation "surface fractals" for the "edges" would be misunderstood. We call it then "edge fractal" or more generally "defect fractal".

The significant amount of this "defected" carbon was confirmed by XPS analysis (fig. 4a, b) while the improved activity nature of the "jaggy edges" rich FLG was shortly investigated by the electrochemical approach in aqueous electrolyte (fig. 4c), both referred to the FLG with straight etched channels. The enhanced contribution of the "non-graphitic" carbon related to the high density of edge defects is reflected by enormous broadening of the C1s XPS peak towards higher binding energies (4a vs. 4b). The full width at half maximum (FWHM) is almost twice (2.8 eV /1.5 eV) with significantly increased contributions from $sp^3$ carbon and oxygen – bound carbon, especially of C-OR and COO type groups (SI).

The activity of these defects/groups is revealed by enhanced electrochemical surface area (ECSA), times fourteen (3.88 vs 0.28 $cm^2$ / 1 $cm^2$ substrate), determined by cyclic voltammetry measurements in aqueous 1M NaOH electrolytes (fig. 4c, SI). For the same mass of deposited material, the double-layer capacitance of "jaggy edges" FLG is thirty times higher (3.3 F/g vs. 0.14 F/g). The obtained values of specific ESCA and capacitance are small due to the π-π stacking of the FLG flakes (and presence of Nafion). The interlayer stacking is a common and well known drawback and might be overpassed by an appropriate formation of 3D structures in the final applications, which is not however the goal of the present work. In turn, the enhancement of ESCA and capacitance in "edges fractal" FLG clearly reveals the presence of active and more hydrophilic carbon, and possibly of the increased porosity. These provide larger interface between FLG and electrolyte, still keeping the charge transport ability through FLG sp$^2$C lattice.

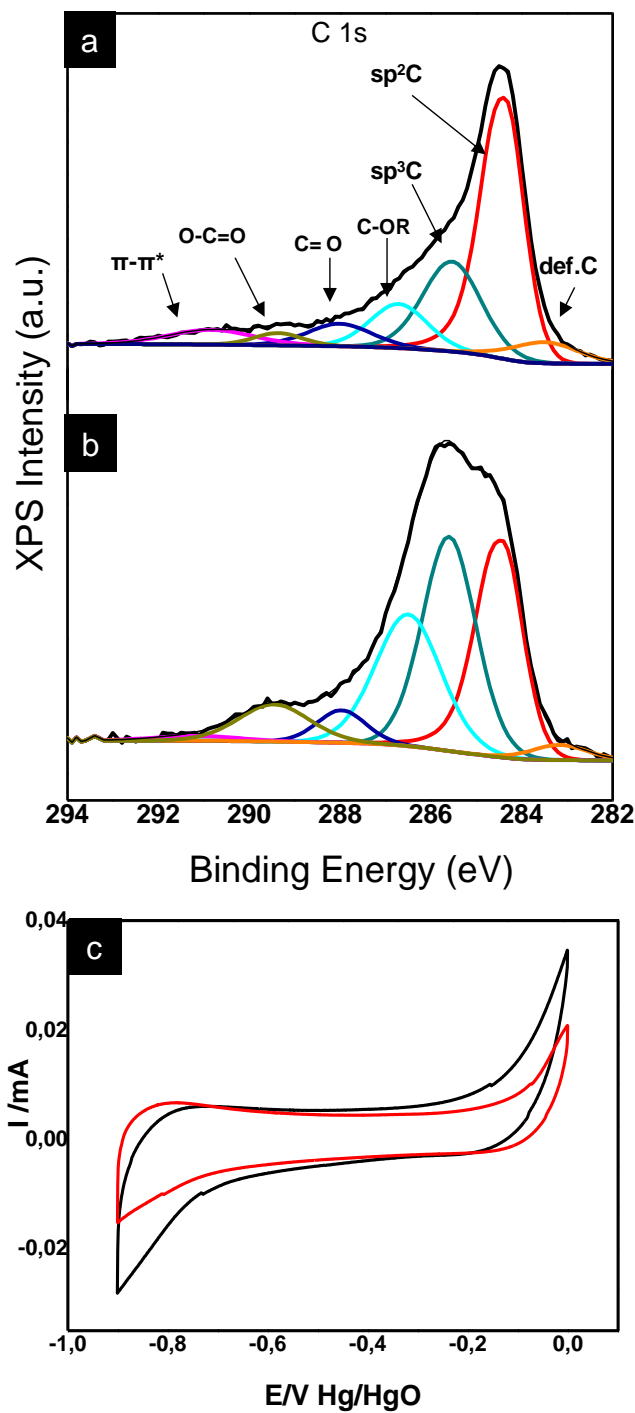

Fig. 4 a,b) Experimental and deconvoluted C1s XPS spectra of (a) FLG with straight etched channels and (b) "jaggy" edges FLG; c) their cyclic voltammetry curves (jagged FLG-black curve)

More than "defect fractal" approach, we would like to herein refer to the fractal nature of graphene it-self and notice that obtained here graphene structures with fractal edges are not far from the model of graphene macro-molecule fractal, which is a benzenoid with rotary symmetry of higher fractal order. The fig. 3c presents such benzenoid with fractal order 4 (n = 4). Let's remind that the fractal configuration will depend on the way the self-similar "circular" units of benzene rings expend in 2D plane. Two different circularly arranged benzenoids structures with a fractal order of 1, 2 and 3 are presented in fig.5 a, b [4, 10]. Clearly, the symmetric structures depicted in fig. 5a reflect Koch snow flake, where triangle corner is replaced by trapeze configuration corresponding to the "arm-chair" conformation of C=C.

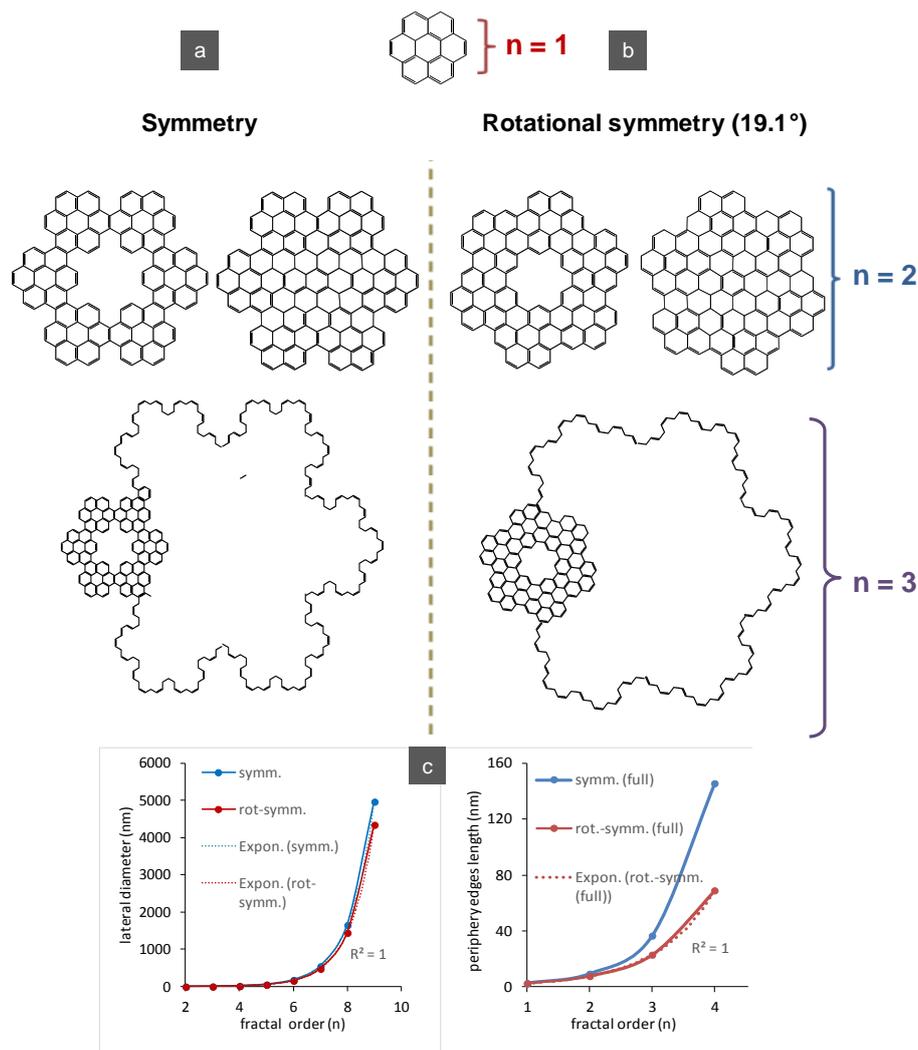

Fig. 5 a, b) schemes of benzenoids structures with increasing fractal order with (a) symmetry (coronene) and (b) rotational symmetry, c) exponential increase of the structures diameter and their periphery edges with fractal order.

The arrangement of the units in the structure b cause the rotational symmetry with the additional counterclockwise rotating angle of 19.1° each time the fractal order increases [4]. Focalizing however on the edges we underline the fact that they increase exponentially and together with the fractal order similarly to diameter of the structures (faster for symmetric structures), and the edges are mixed. Funnily, if we consider the fractal approach for graphene planar lattice and its related properties, the reasoning of dimension reduction from 2D is right; in the case of edge fractal, the 1D dimension increases. The "defect fractal" includes also the holes in the model macromolecules, which are formed by "the luck" of benzenoids units in the plane having as well nonlinearly increasing edges as illustrated in the in fig. 5 and in SI. The illustrated here schema of benzenoids, coroenenes including (fig.5), are up to now the theoretical "idealistic" relatively small structures and their preparation through a total organic synthesis would be difficult but easily controlled. Other way to achieve such structures at high scale (omitting a laser or lithography patterning) would be a well controlled catalytic etching. The latter route seems also be more realistic for the structures with higher size (higher fractal order).

In order to explain the exact process of the edges etching and to well control the final structures, the additional efforts needs to be undertaken and the supporting in-*situ* TEM and XPS observations are considered for this purpose in future. The process to get "jagged" but still sharp and well defined edges is complex and both, oxygen pretreatment and hydrogen are necessary. At this moment it is difficult to determine the impact of the enhanced oxygenation degree of FLG once the temperature is increased and hydrogen is supplied. Regardless the oxidation with atomic oxygen and assuming that C-metal interactions are important factor, different phenomena can take place during the etching. These are NPs coalescence, splitting, melting, faceting and structure reorganization [3, 11, 12], which, as chemistry of the NPs-FLG hybrid changes under etching process (iron, iron oxides and carbide phase) [13], will depend not only on the etching conditions but also on the preparation and pretreatments of the hybrid. We suggest that the pre-oxidized edges of FLG have increased affinity towards the NPs inducing a possible splitting of the latter and allowing for fine-grained jaggy etching due to the small size NPs (or their *quasi*-melt phase), while the etching of more defected (active) carbon it-self happens easier. From the other hand, the faceting is expected especially for larger NPs as observed in *"post-mortem"* catalyst and has an impact on the edges structure in general.

The presented concept of etching to edge fractal structures is different from the one reported for the in-situ CVD synthesized graphene via Cu catalyst, where either hexagonal structures or sharp thin highly etched branched structures were obtained [14, 15]. In herein approach, the significant area of honeycomb structure remains robust while the edges are highly etched. This leads to the bulk materials with still significant planar surface, related high mechanical resistance and platform for charge transport, but also great density of nonlinearly etched edges active sites. The more controlled etching would entail not only controlled defect density but also surface area and porosity (size, volume, distribution). Such tailoring and activation of large and bulky graphene structures are benefic for many applications including the electrodes (supercapacitors, fuel cells), where enhanced (electro)catalytic activity, appropriate porosity, surface area and electrode-electrolyte interface interactions, but also sufficient conductive platform for efficient charge transport, are reached. Number of other fields would gain from controlled fractal edges tailoring, not to mention the electronic, optical, sensing or magnetic properties, especially in the case of well defined smaller fractal order structures. The efforts for their preparation and theoretical calculations should be then made. The in-*situ* CVD-fractal etching of the individual graphene sheets would seem to be easier way for the etching control that the etching of "bulk" graphene, however, the in-*situ* CVD-fractal etching over catalytic substrate has similar advantages and drawbacks as CVD synthesis of graphene. Despite the easier control of the final patterns, the use it for "bulk" applications is strongly scale limited along with the substrate transfer issue, which would be much harder in the case of etched structures.

**Autor contributions:** I. Janowska: principal author, project investigator and coordinator, writing; M. Lafjah: initial experiment, V. Papaefthymiou: XPS analysis, S. Pronkin: electrochemical analysis, C. Ulhaq-Bouillet: TEM microscopy.

**Supplementary Information:** The methods are available